\documentclass[aps,twocolumn,amsmath,amssymb]{revtex4}
\usepackage{bm}

\usepackage[dvips]{graphicx}
\usepackage{dcolumn}
\usepackage[mathcal]{euscript}
\usepackage[usenames,dvipsnames]{color}

\def\beq{\begin{equation}}
\def\eeq{\end{equation}}

\begin{document}
\bibliographystyle{apsrev}

\title{Microtrap arrays on magnetic film atom chips for quantum information science}

\author{V.Y.F. Leung}
\author{A. Tauschinsky}
\author{N.J. van Druten}
\author{R.J.C. Spreeuw}
\email{r.j.c.spreeuw@uva.nl}
\affiliation{Van der Waals-Zeeman Institute, University of Amsterdam, \\
Science Park 904, PO Box 94485, 1090 GL Amsterdam, The Netherlands}
\homepage{http://www.science.uva.nl/research/aplp/}

\date{\today}

\begin{abstract} 

We present two different strategies for developing a quantum information science platform, based on our experimental results with magnetic microtrap arrays on a magnetic-film atom chip.  The first strategy aims for mesoscopic ensemble qubits in a lattice of $\sim 5\,\mu$m period, so that qubits can be individually addressed and interactions can be mediated by Rydberg excitations. The second strategy aims for direct quantum simulators using sub-optical lattices of $\sim 100$~nm period. These would allow the realization of condensed matter inspired quantum many-body systems, such as Hubbard models in new parameter regimes. The two approaches raise quite different issues, some of which are identified and discussed.

\end{abstract}



\maketitle

\section{Introduction}

Atom chips are a promising technology for developing a novel QI science platform \cite{ForZim07}. They combine the best of two worlds, on the one hand neutral atoms with their weak coupling to the environment and concomitant long coherence times \cite{TreHomRei04,DeuRamRos10}, and on the other hand the compactness and large-scale integration possibilities of solid-state like systems.  

Here we present our experiments using an array of individually addressable magnetic microtraps defined on a magnetic-film atom chip \cite{GerWhiSpr07,WhiGerSpr09}.  We discuss the scaling issues, and prospects to develop a quantum simulator, which for the first time is becoming technologically realizable.  A series of theoretical proposals \cite{JakCirLuk00, LukFleZol01, WeiMulBuc10} has suggested that experimental techniques available today can be synthesized into an universal device capable of mimicking the behavior of complex, many-spin systems.

Typically, simulating quantum many-body physics has required a lattice structure into which individual particles, whether neutral atoms, ions, or electrons, can be loaded and then coherently controlled.  Analog quantum simulators can be used to find the ground state of interacting quantum many-body systems such as Hubbard models. Coherent control can also be used to prepare the simulator in an analogous initial state, after which its subsequent dynamic evolution naturally mimics that of the system of interest.  Of greater versatility is the digital quantum simulator, in which quantum gates are operated to produce unitary time evolutions. This is also known as an universal quantum computer \cite{BulNor09}.

Both types of simulators can be implemented with neutral atoms in a lattice.  For atom chip devices, lattices formed by magnetic potentials offer some promising technical prospects compared to other approaches, such as optical lattices. Arbitrary lattice configurations and spacings can be implemented without the need to consider the wavelength of the light or optical access close to the chip surface.  This becomes particularly critical when considering the exciting regime of sub-optical wavelength lattice spacings on the order of 100 nm.

As we have recently demonstrated a working shift register on a magnetic lattice of $20\;\mu$m spacing \cite{WhiGerSpr09}, the major questions we are addressing in this article can be summarized as ``the physics of scaling down''.  We present our estimates for the required specifications of a magnetic-lattice-based register at $5\;\mu$m and, ultimately, 100 nm.  We will be considering issues in the context of both the implementation of a digital quantum simulator composed of Rydberg gates and a true analog quantum simulator at a lattice spacing of 100 nm.

For the micron-scale array of mesoscopic ensemble qubits, we describe our results on three-body decay leading to sub-Poisson atom number fluctuations \cite{WhiOckSpr10}.  We also discuss our progress on sensitive detection of small atom numbers and the investigation of Rydberg excitation close to the chip surface.  First experiments on Rydberg atoms on chips have already been performed, in which we measured surface-based electric fields using electromagnetically-induced transparency \cite{TauThiSpr10}. 

The sub-optical scale lattices can potentially yield a novel realization of a Hubbard model system that can open up new parameter regimes. We discuss the scaling of the Hubbard model parameters with our atom chip in mind.  We calculate the influence of the Van der Waals surface attraction and show that the magnetic trapping potential is strong enough to overcome this. We also calculate rates for tunneling into the surface, and show that for practical cases tunneling into the surface can be neglected.  Finally, we estimate trap loss rates due to Johnson-noise-driven spin flips and explore ways to mitigate this effect. 

A non-trivial technical challenge which accompanies increasing minaturization is the question of effective detection of closely spaced atoms on the single-particle level.  Although much of our research efforts have been towards engineering magnetic trapping potentials, we shall briefly discuss an approach to tailor {\em electric} fields, based on ideas inspired by earlier work on field emission from arrays of silicon tips \cite{TeeVeeKru05} and from individual carbon nanotubes \cite{JonDru03}. This offers several opportunities, in particular for detection based on site-selective field ionization on a magnetic atom chip.

\section{Microtrap array on a magnetic-film atom chip}

\begin{figure}[h]
\centering
\includegraphics[width=0.5\columnwidth]{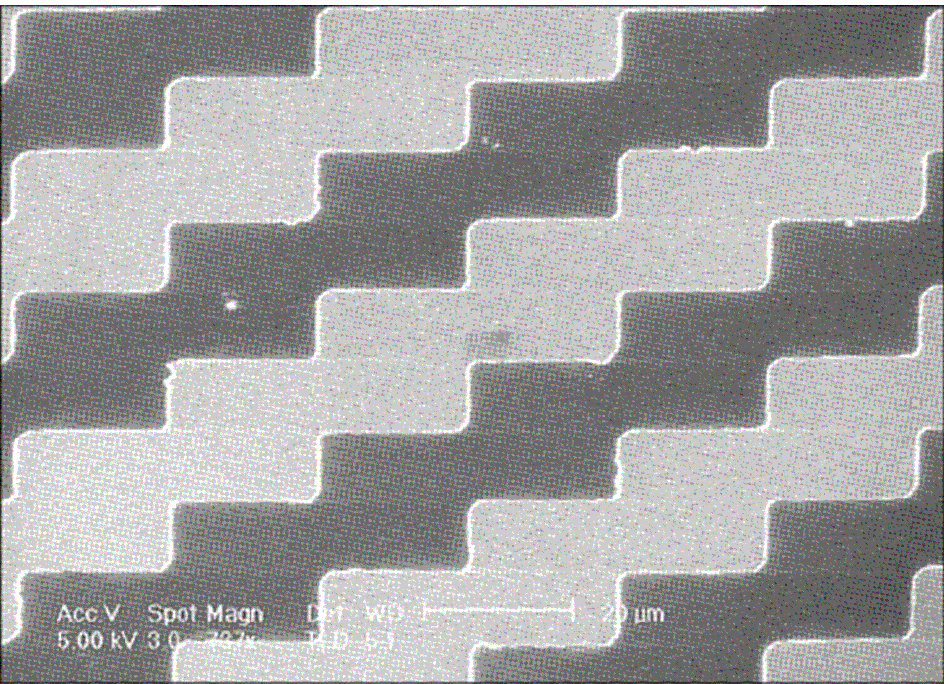}\\
\vspace{6mm}
\includegraphics[width=0.9\columnwidth]{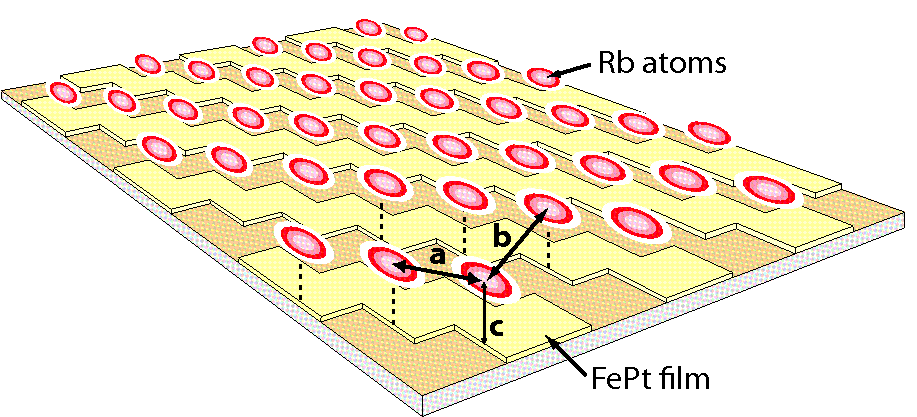}
\caption{Upper: scanning electron microscopy image of the patterned magnetic film.  The light grey areas correspond to the FePt pattern and the darker regions are the Si substrate. The size of the horizontal edges is 20~$\mu$m. Lower: schematic representation of the microtrap array on the magnetic-film atom chip. Small ultracold ($\sim\mu$K) clouds of Rb atoms are trapped above edges of the magnetized FePt-film pattern, at $c \approx 10\;\mu$m from the surface. Lattice spacings are $a = 22\;\mu$m and $b = 36\;\mu$m. }
\label{fig:latticeschem}
\end{figure}

We briefly review our previously reported results on an array of microscopic Ioffe-Pritchard type magnetic traps, defined by patterning a permanently magnetized film on a chip \cite{GerWhiSpr07,WhiGerSpr09}.
The film with a thickness of 300 nm consists of FePt with a remanent magnetization of 670~kA/m oriented perpendicular to its surface, and is deposited on a 300~$\mu$m-thick silicon substrate. 
Such high remanent magnetization is an important advantage of FePt, in addition to its high coercivity, a high Curie temperature that allows mild baking of the vacuum system, and its corrosion resistance. FePt is also very suitable for patterning via conventional lithographic techniques.  Transfer of the pattern was made by UV lithography and plasma etching, resulting in the pattern shown in Fig.~\ref{fig:latticeschem}, after which the chip is coated with a reflecting gold layer.  This process yields a two-dimensional array of 1250 traps/mm$^2$.  Each trap hovers 10--15~$\mu$m above the chip surface, see also Fig.~\ref{fig:latticeschem}; the traps are tightly confining, with trapping frequencies of $\sim 10$~kHz for $^{87}$Rb atoms in the $|F = m_{F} = 2\rangle$ state.

\begin{figure}[ht]
\center
\includegraphics[width=\columnwidth]{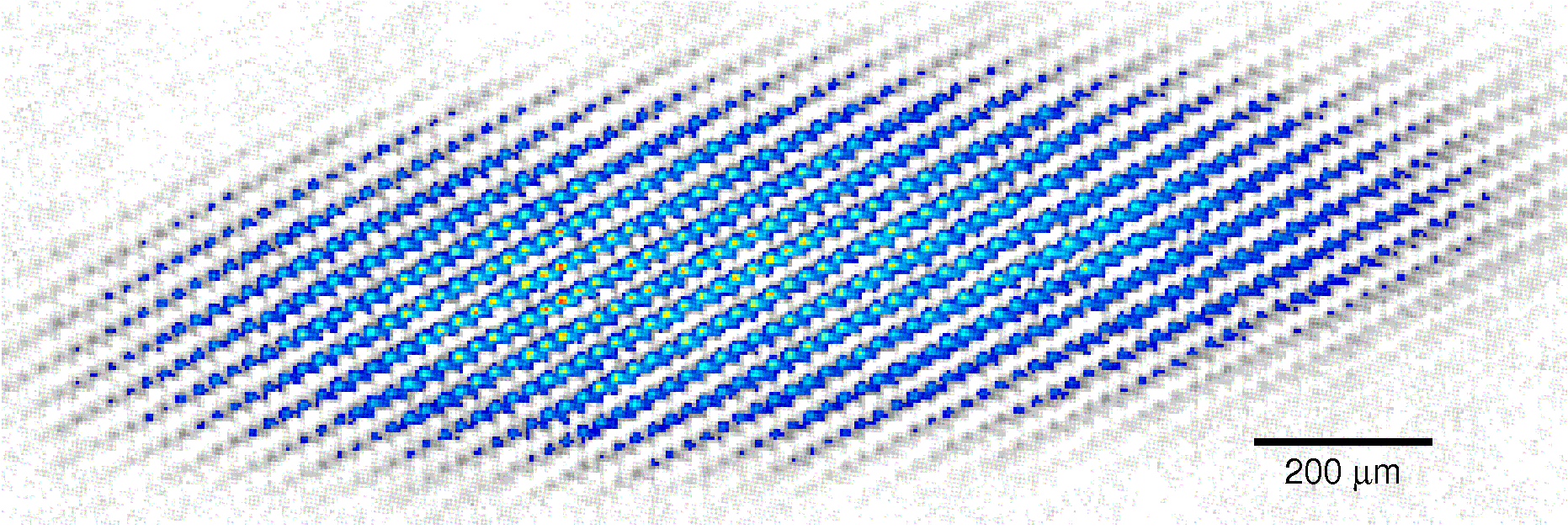}
\caption{Absorption image of the loaded lattice, showing $\sim500$ traps loaded with 200-2500 atoms each \cite{WhiGerSpr09}.}
\label{fig:latticeimage}
\end{figure}

The  chip is mounted on a copper surface with an integrated Z-wire, which is used for loading in conjunction with an external bias field.  Radio frequency evaporative cooling is used to lower the temperature of the atoms to quantum degeneracy in the micro-Kelvin regime.
Figure~\ref{fig:latticeimage} shows an averaged absorption image of the loaded magnetic lattice. Here, $\sim 500$ traps are loaded with between $200$ and $2500$ atoms each. Individual traps are resolved, with an optical
resolution of $\approx 7.5~\mu$m (Rayleigh criterion), enabling individual detection of the number of atoms in each ensemble. Our microtrap array of near-degenerate mesoscopic atomic ensembles bridges the gap between optical lattices with a single or few atoms per site on the one end, and single macroscopic Bose-Einstein condensates with $10^3-10^8$ atoms on the other end. 

\begin{figure}[ht]
\center
\includegraphics[width=\columnwidth]{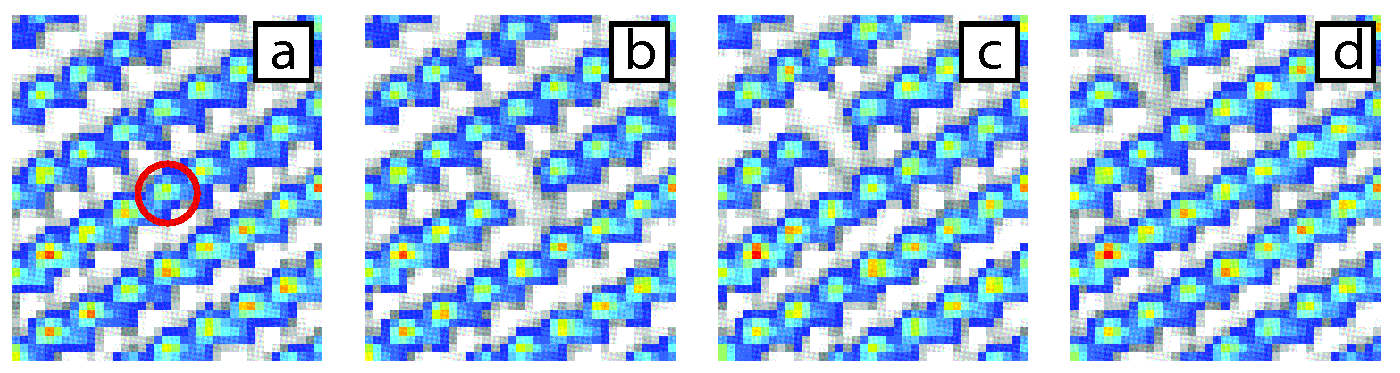}
\caption{Optical addressing of single lattice sites and atomic shift register. (a) Close up absorption image of the central region of the lattice. The trap at the focus point of the addressing laser is circled in red. (b) A single site has been emptied by a $1$~ms laser pulse. (c-d) The entire lattice is shifted two cycles by rotating the external magnetic bias field \cite{WhiGerSpr09}.}
\label{fig:manipulation}
\end{figure}

It is possible to address individual sites and selectively empty one site using a focused laser pulse.  The optical pumping transition $F = 2 \rightarrow F' = 2$ of the D$_2$ line is used to  flip the spin of the atoms in a selected site into a magnetically anti-trapped state.  This can be combined with parallel transport of atoms along the array, which has also been demonstrated \cite{WhiGerSpr09}.  External magnetic fields were dynamically varied to shift the potential minima of the micro-traps across the surface of the chip.    Figure~\ref{fig:manipulation} shows shifting over two lattice periods, using a vacancy made by the addressing laser as a marker. During the shifting, the empty sites do not refill, demonstrating there is no interaction between neighbouring sites.  Atoms at $\approx$~10~$\mu$K were transported 360~$\mu$m using this method, without significant additional heating.

The magnetic microtrap array is an excellent starting point to develop a scalable quantum information science platform. In the remainder of the paper we discuss some aspects of two different strategies that we are pursuing to achieve this goal. 

The first involves the redesign and scaling down of the array to arrive at square and/or triangular lattices of $\sim 5\;\mu$m period. This would allow the definition of an ensemble qubit at each lattice site, for which the transition between the logical states $|0\rangle$, $|1\rangle$ is a collective, single-atom excitation to another ground hyperfine state. This approach is based on proposals relying on dipole blockade between highly excited Rydberg atoms \cite{JakCirLuk00,LukFleZol01,BriMolSaf07,MulLesZol09,WeiMulBuc10}. 

The second strategy involves scaling down the array to sub-optical length scales of $\sim 100\;$nm period. This may open up new parameter regimes in Hubbard model physics that are as yet out of reach for optical lattices, allowing the realization of direct quantum simulators of, for example, condensed-matter-inspired systems. In this approach the atoms will be trapped much closer to the surface. This makes the use of Rydberg atoms unrealistic and makes it necessary to carefully consider the atom-surface interaction.

\section{Mesoscopic ensembles for quantum information}

The use of mesoscopic ensembles for quantum information hinges on the use of strong, long-range interactions among highly excited Rydberg atoms to orchestrate switchable gates \cite{JakCirLuk00,LukFleZol01,BriMolSaf07}. In the dipole blockade effect, the strong electric dipole moment or polarizability of an atom excited to a Rydberg state shifts the energy levels of nearby Rydberg atoms, prohibiting the excitation of more than one atom to the Rydberg state within one so-called blockade radius. 
The strong, long-range, and switchable interaction  fuels the interest in Rydberg atoms for implementing quantum gate protocols and multi-particle entanglement \cite{MulLesZol09,SafMol09,WeiMulBuc10}.  

Dipole blockade has been proposed as a way to define ensemble qubits and to create entangled mesoscopic ensembles. So far the dipole blockade has been observed either in clouds much larger than a blockade radius or among separate single atoms \cite{HeiRaiPfa08,UrbJohSaf09,GaeMirGra09}.
Recent experiments have demonstrated dipole blockade between single atoms up to ~10$\;\mu$m apart and CNOT gate operation between two individually addressed neutral atoms.  
Resonant energy transfer has been demonstrated over 40$\;\mu$m \cite{DitKoeLin08}. 
For comparison, an atom in the ground vibrational level of a 10~kHz microtrap is confined to less than 100~nm. Even an ensemble at a temperature of $T=2\;\mu$K would have a rms size of only 200~nm. Thus the separation between mesoscopic ensembles in strongly confining microtraps are well within the Rydberg interaction radius, making them a promising candidate qubit. 

In ensembles, qubits could be encoded in collective excitations produced using intratrap dipole blockade. We consider two trappable hyperfine ground states, for example the $|F,m_F\rangle = |1,-1\rangle\equiv |A\rangle$ and $|2,1\rangle\equiv |B\rangle$ states in $^{87}$Rb, which have nearly the same magnetic moment. At the special magnetic field value of $3.23$~G these two states serve as ``clock states'' with vanishing first order differential Zeeman shift \cite{TreHomRei04}. The logical qubit states in an ensemble of $N$ atoms could be defined as follows: $|0\rangle\equiv |A\rangle^{\otimes N}$ and $|1\rangle\equiv |A\rangle^{\otimes (N-1)}|B\rangle$. Thus the logical $|1\rangle$ is a single, collective hyperfine ground state excitation, which can be accessed via an intermediate collective Rydberg excitation if the entire ensemble is within one blockade radius. 
The strong confinement in magnetic microtraps brings us in this interesting regime where the dipole blockade allows the creation of a single collective Rydberg excitation $ |A\rangle^{\otimes (N-1)}|r\rangle$, which can be mapped back on $|1\rangle$ by a $\pi$ pulse \cite{WilGaeBro10,IseUrbSaf10}.

The prospect of using controlled Rydberg interactions also among neighboring microtraps have led us to redesign the array. 
Where the previous design had a parallelogram unit cell with lattice spacings of 22~$\mu$m and 36~$\mu$m, respectively (Fig.~\ref{fig:latticeschem}), the new design features a square and a triangular lattice both with a period of 5~$\mu$m in all directions. Equal lattice spacings are a requirement for truly 2D nearest-neighbour interactions. The 5~$\mu$m spacing is chosen to balance strong Rydberg dipole-dipole interactions with sufficient optical resolution to resolve individual traps with in-vacuum optics.
Thus the high trap frequencies ($>10$~kHz) and relatively small lattice constants of approximately 5~$\mu$m will allow us to investigate both intra- and inter-trap interactions. 
The new magnetic lattice patterns have been designed using a numerical optimization algorithm \cite{SchLeiWhi10}. The resulting magnetization pattern for the square lattice is shown in Fig.~\ref{square1d}(a) in a downscaled version.  

The approaches sketched here raise several issues that we will discuss in more detail.


\subsection{Sub-Poisson atom number fluctuations}

In collective excitation schemes, an ensemble containing $N$ atoms would benefit from a $\sqrt{N}$ enhancement in Rabi frequency, increasing the number of operations achievable within the coherence time \cite{LukFleZol01}. However, the $N$-dependence of the Rabi frequency adversely affects the fidelity of operations if the number of atoms fluctuates from shot to shot and/or from site to site. Intrinsic atom number fluctuations would typically lead to a Poisson distribution at best; however, the fluctuations can be reduced to sub-Poissonian levels by three-body losses.  In our case we find that three-body loss is the dominant trap loss mechanism due to the high trap frequencies and corresponding high densities \cite{WhiOckSpr10}.

We find theoretically and experimentally that the normalized atom number variance, or Fano factor, $F=\langle N^2\rangle/\langle N\rangle$ is reduced to the value 0.53(22), significantly below the value of 1 corresponding to Poisson statistics. Furthermore, the memory of any initial fluctuations is quicky erased. The Fano factor decays as the fifth power of the remaining fractional atom number $\eta$, 
\begin{equation}
F(\eta) = \frac{3}{5} + \eta^5\left(F_0 - \frac{3}{ 5}\right).
\end{equation}
This means in practice that once half the atoms are lost, the Fano factor is usually below 1 and already close to its asymptotic value of 3/5. 

The occurrence of sub-Poissonian atom number fluctuations can be seen as an atomic analog to intensity squeezing in optics. By trapping a large number of dense mesoscopic ensembles in a lattice of microtraps which undergo rapid three-body decay, it was shown that three-body loss can be used to prepare small and well-defined numbers of atoms in each trap. Through sensitive absorption imaging the shot-to-shot distribution of atom numbers was measured and found to have sub-Poissonian statistics for between $50$ and $300$ atoms per trap, in good agreement with a model for stochastic three-body loss which takes into account the fluctuations.

\subsection{Rydberg-surface interaction}

Little has so far been known about the influence of the nearby surface on Rydberg atoms created on an atom chip \cite{AndHarMes88,SanSukHar92}. We investigated such surface effects \cite{TauThiSpr10} using excited-state electromagnetically induced transparency (EIT) \cite{MohJacAda07,WeaPriAda08} and extending this technique to obtain EIT spectra at distances between $\approx 20 - 200 \mu$m from the chip surface. The position and width of the narrow transmission resonance reflect the energy and lifetime of the Rydberg state under investigation and provide a sensitive probe of the atom-surface interaction. 

We find strong shifts of the Rydberg states of several 10s of MHz for principal quantum numbers $n \approx 30$ at these distances. The magnitude and sign of the shift are determined by the polarizability of the Rydberg state; they are caused by a dipolar electric field created by a patch of Rb adsorbed on the chip which has been deposited there over many experimental cycles during regular operation of the experiment \cite{ObrWilCor07}. We also find that there is no broadening of the Rydberg resonances above the level given by the linewidth of the lasers used in the experiment, either by this patch of adatoms or by other sources.

The surface effects due to adatoms could be prevented in future experiments by incorporating a magnetic field gradient to push the atoms away from the surface at the end of each experimental cycle, or the adatoms could be removed using e.g. light induced desorption \cite{SteBenSla10}. Furthermore, use of other coating materials on the chip surface could decrease the dipole moment of the adatoms or increase the desorption rate. Finally, one could adjust the frequency of the coupling laser to compensate for any residual effects. 

The absence of any broadening and the proposed measures for dealing with the level shifts caused by adatoms lead to the conclusion that surface fields do not cause any serious obstacles to creating Rydberg atoms at distances of order 10~$\mu$m from the surface of an atom chip.

\subsection{Sensitive detection of small atom numbers}

An important technical challenge lies in the state-resolved detection of either a single atom or a single collective hyperfine excitation of an ensemble. 
The direct detection of atoms  in their Rydberg state would require additional technology such as ion detection, etc.  Although the Rydberg atom can be mapped back on a detectable ground state such as the $|F=2\rangle$ by a $\pi$-pulse in the regime of active dipole blockade, this would still require the capability to detect at the single-atom level.

Our group has recently demonstrated sensitive detection of small ($\sim 10$) numbers of atoms by reflective absorption imaging, using advanced image processing \cite{OckTauWhi10}. First, a fringe removal algorithm was used to reduce imaging noise to the fundamental photon-shot-noise level. 
A maximum-likelihood estimator (MLE) was then used for optimal atom-number estimation. The MLE is optimal in the sense that it achieves the Cram\'er-Rao bound: a lower bound for the variance of any parameter estimate, independent of the exact procedure used to extract
the information. In particular the variance in the estimated atom number was shown to be lower than one would obtain from a simple integration over a region of interest in the image. 
Both the detection sensitivity and the spatial resolution that were achieved were limited by the low (0.1) numerical aperture (NA) of the imaging optics.  Increasing the NA is straightforward by mounting a high NA lens inside the vacuum chamber.  Our estimate shows  that  single-atom, single-shot readout sensitivity using absorption imaging is entirely realistic.

This capability will be important, apart from the state-selective readout of ensemble qubits, to verify the phenomenon of dipole blockade.
Two important milestones will be to detect the blockade by one Rydberg atom of further excitation to the Rydberg level of all other atoms in the same potential well, followed by the detection of  blockade of Rydberg excitation in adjacent wells.

In addition to reaching single-atom sensitivity in the readout there is a range of possibilities to extend our detection capabilities.  One powerful option is to make use of the lattice periodicity. A simple averaging over lattice sites will reduce the uncertainty in the average atom number per site. More detailed information is available however, correlating the imaging data of neighboring sites.  The use of correlation  functions has already allowed us to separate various noise sources in our measurements of sub-Poissonian atom number fluctuations. Whereas averaging can show that the average number of collective excitations within one site is 1, correlation functions can be to used show that double occupancy is suppressed \cite{GreRegJin05}.  In a similar fashion correlation functions should boost the sensitivity to detect entanglement between neighboring sites. 

\subsection{Site-selective electric fields} \label{electricfields}

An interesting alternative to optical single-particle detection is the use of sharp tips for ionization-based detection.  Arrays of such tips have been fabricated in silicon for field emission purposes \cite{TeeVeeKru05}.  Such arrays could be integrated on the atom chip and are site-selective by nature. In combination with the shift register operation that we have demonstrated before \cite{WhiGerSpr09}, a linear array of tips would be sufficient to read out a two-dimensional lattice. 

An externally applied homogeneous electric field is concentrated at the tip of a protrusion;
 this is the ``lightning rod'' effect. Therefore, structuring the surface topology of a conductor can be used to tailor the local electric field.  At a distance $r$ from the center of a tip of a single thin, long post (height $h$), the external field is enhanced by a factor $\gamma\approx h/r$ \cite{ForEdgVal03}. 

For the magnetic lattices of several micrometer periodicity considered in this section, it should be possible to achieve useful site-selective electric fields with conventional semiconductor technology.  The arrays of silicon field emitters produced and characterized in Ref.~\cite{TeeVeeKru05} are very promising in this regard. With external electric fields of a few kV/mm, field emission from the tip of 
$\approx 5~\mu$m-high cone-like protrusion was achieved, corresponding to a field enhancement factor $\beta>10^3$.  This height is well matched to the present magnetic-film structures. 

Using silicon technology, a chip can be fabricated with surface protrusions at selected sites.  To maintain conceptual simplicity, the protrusions can be  located at positions where the magnetic film is absent, to avoid large changes in the local magnetic-field structure.  Application of a homogeneous external electric field (generated by voltage difference between the chip and a remote counterelectrode) will then lead to substantially enhanced  electric fields near the tip of the protrusions.  Because the field is static, only modest conductivity of the surface is needed to achieve the field enhancement, compatible with the present FePt magnetic-film atom chips.

The highly inhomogeneous electric field will exert a force on the atoms (proportional to their polarizability) directed towards the tip. For a strong field and narrow tip, the atoms will be accelerated and field-ionized when the field is sufficiently high. The resulting ions or electrons are further accelerated by the electric field towards the surface and counterelectrode.  This can be detected at  the single-particle level (e.g. by using a channeltron or multichannel-plate detector at the counterelectrode), and thus appears promising as a non-optical site-selective single-particle detector.  


\section{Scaling down to sub-optical period lattices}

Magnetic trap arrays provide a range of opportunities that experiments so far have only 
started to explore. 
They can provide 
versatile, tunable model
systems for the study of strongly interacting quantum many-body systems, in potentially  
new parameter regimes beyond what is accessible with optical lattices. In this section we will investigate scaling down a two-dimensional trap array to sub-optical dimensions, with a lattice parameter of 100~nm. 

This would yield a new implementation of the Hubbard model, the prevailing paradigm for quantum gases in optical lattices. Scaling down to substantially smaller lattice parameters will increase all relevant energy scales, opening up new regimes with qualitatively new physical phenomena. All key parameters, including the tunneling rate and on-site interaction can be dynamically tuned. 

Virtually unlimited design freedom allows for, among others, square, triangular, hexagonal (graphene-like), and Kagome lattice geometries. In addition, controlled amounts of disorder and designer defects can be built in.  

In this section we consider some of the physics issues of this nanoscale scenario.

\subsection{Scaling and tunability of Hubbard model parameters}

The Hubbard model is widely used to describe quantum gases in optical lattices \cite{GreManBlo02}. It is characterized by the tunneling rate $J$ (also called $t$) and on-site interaction $U$. These energy scales can be  naturally expressed in units of the lattice recoil, $E_R=(\pi\hbar)^2/2md^2$, where $d$ is the lattice constant \cite{GerWidBlo05}. A typical value for optical lattice experiments to date has been $d=425$~nm. Lattices with period $d=100$~nm or even smaller would allow a realization of the Hubbard model in a novel regime that is as yet not accessible by optical lattices. Tunneling rates will become very large, as will the on-site interaction.  This may bring energy scales such as the superexchange rate $J^2/U$, marking the regime of quantum magnetism, within easier reach.

The result of reducing the lattice parameter from 425 to 100 nm is an increase of the recoil
energy $E_R$ from 153 nK to 2.75 $\mu$K. In Table~\ref{table:scalingdown} we compare a few other energy scales for
the two situations, based on scaling expressions for $U$ and $J$ given in Ref.\ \cite{HoZho07} for sinusoidal potentials. As a
benchmark we calculate $U$, $J$, and $J^2/U$ for the value of the lattice depth $V_0/E_R$ that
corresponds to the superfluid to Mott insulator transition. 
For a two-dimensional square lattice the transition was found experimentally at $J/U=0.06$ by Spielman {\em et al.\ }\cite{SpiPhiPor07}. For $d=100$~nm this corresponds to $V_0/E_R=6.2$.
Table~\ref{table:scalingdown} shows that all energy scales are raised by one to two orders of magnitude, thus relaxing the extreme
requirements as posed by the optical lattice parameters. 
The quantity $J^{2}/U$ is the energy scale
associated with superexchange (quantum magnetism). It can be seen that scaling down the lattice yields a value for $J^{2}/U$ of 8~nK, well above the lowest reported temperature \cite{LeaPasKet03}.

\begin{table}[ht]
\centering
\begin{tabular}{c c c}
\hline 
$d$ & 425 nm & 100 nm \\ [0.5ex]

\hline \hline 
$V_0/E_R$ & 10.4 & 6.2\\
\hline 

$E_R$ & 153 nK & 2.75 $\mu$K \\
$U$ & 46 nK & 2.2 $\mu$K \\
$J$ & 2.7 nK & 135 nK \\
$J^2/U$ & 0.16 nK & 8.1 nK  \\ [0.5ex]

\hline
\end{tabular}
\caption{Comparison of energy scales for an optical lattice (period $d=425$~nm)
and a magnetic lattice with $d=100$~nm. The value for the lattice depth
$V_0/E_R$ has been chosen such that $U/4J=4.2$, corresponding to the Mott insulator
transition for a 2D square lattice as measured in Ref.~\cite{SpiPhiPor07}. }
\label{table:scalingdown}
\end{table}

A lattice period of 100 nm will thus bring us into an exciting new regime. The trap-to-surface distance will be scaled down in proportion, because the magnetic field of any Fourier component in the surface magnetization with wave vector $\bm k$ decays exponentially as $B \sim\exp(-kz)$. Answering the question of how far we can ultimately scale down will therefore depend crucially on our understanding of the interaction with the surface. The value of 100 nm appears challenging but feasible. We briefly discuss the orders of magnitude of a few different aspects of the surface interaction that will play a role in our case.

\begin{figure}%
\center
\includegraphics[width=0.8\columnwidth]{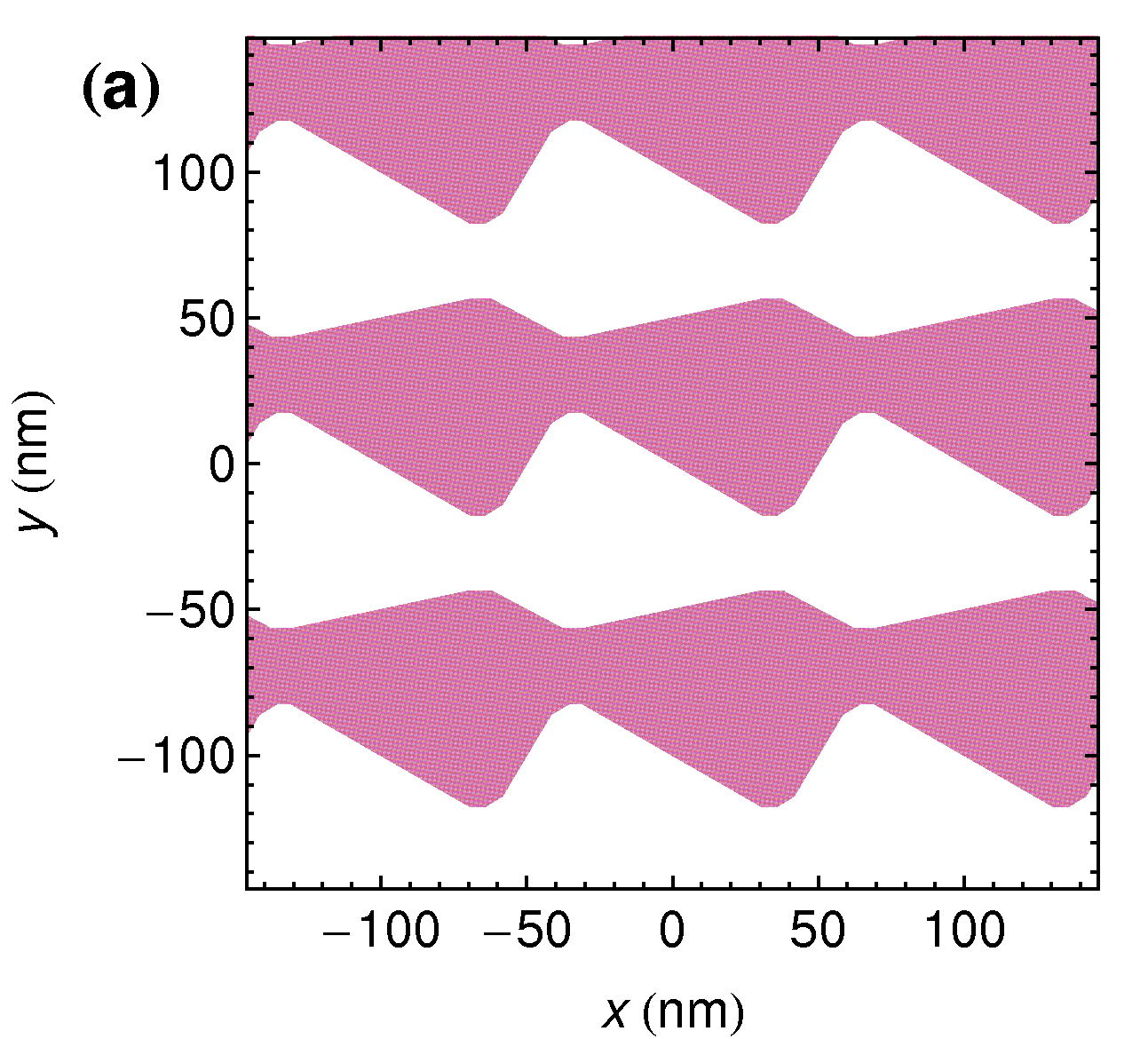}\\
\includegraphics[width=0.8\columnwidth]{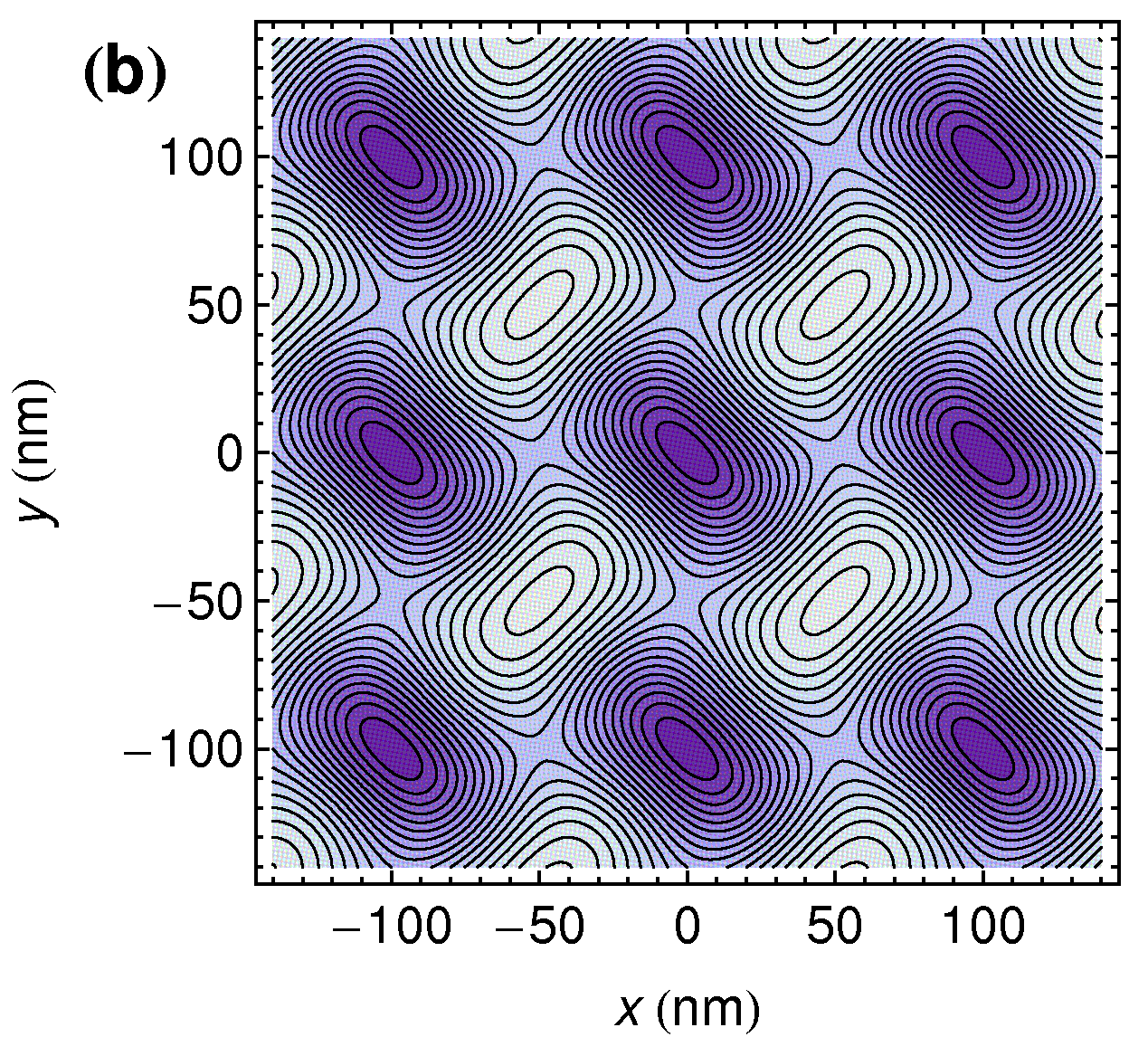}\\
\includegraphics[width=0.8\columnwidth]{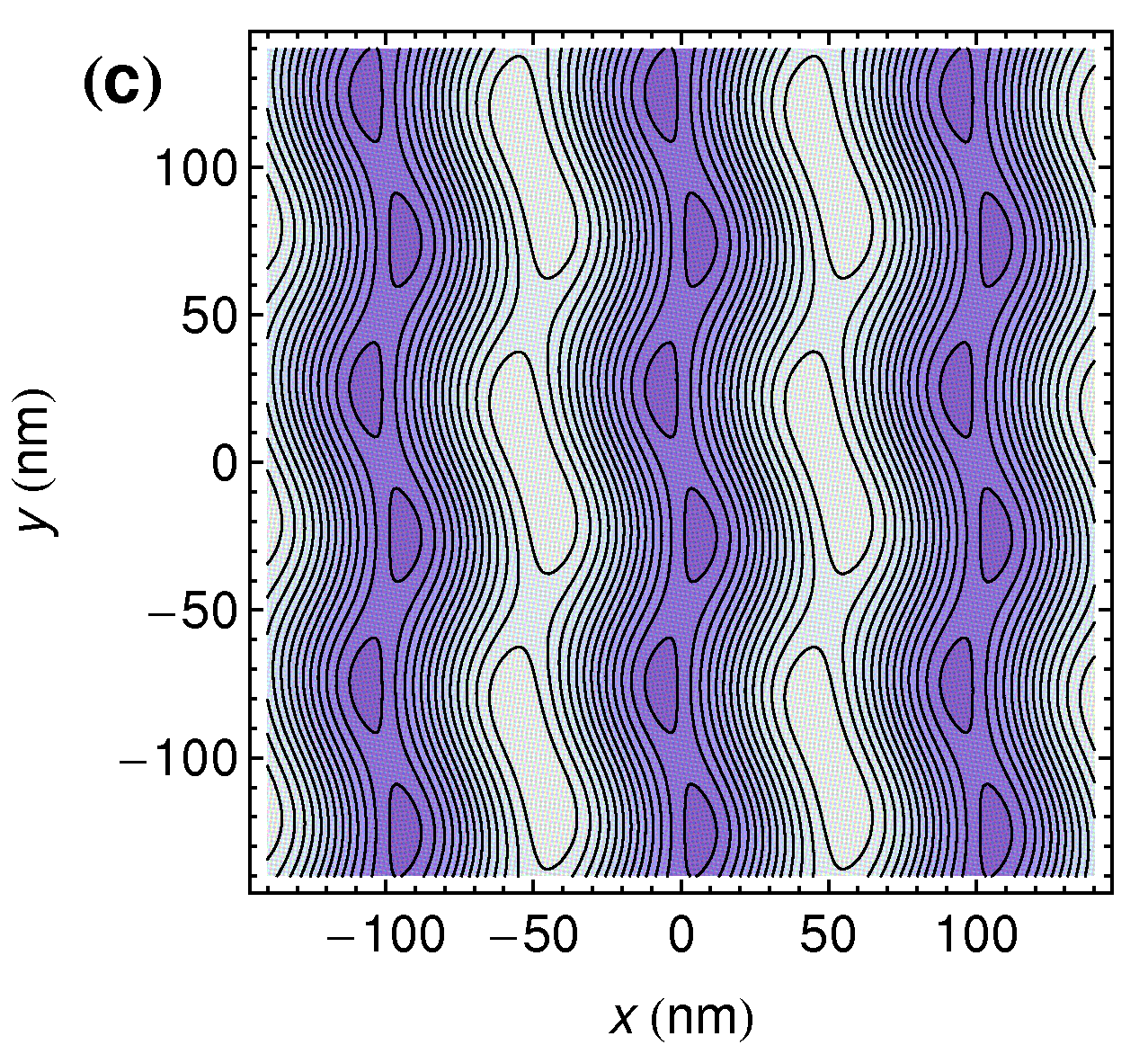}%
\caption{(a) Optimized magnetization pattern for a square lattice. (b) Square lattice, biased for symmetric barriers and trapping minima at $z=100$~nm. Drawn in the plane of the trapping minima. Contours are lines of equal $B$, at 0.5~G intervals. Darker shades indicate lower fields, i.e.\ lower potential. (c) Same pattern biased to yield 1D channels along $y$, again drawn in the plane of the trapping minima with 0.5~G contour spacing.}%
\label{square1d}%
\end{figure}

\subsection{Optimized design}

The best design for a magnetic film pattern which would result in a microtrap lattice with specific trapping parameters is not always self-evident, most often relying on experience from numerous trials.  A fast linear programming algorithm has been developed to automatically generate magnetization patterns which provide optimal atom confinement while respecting desired lattice symmetries and trap parameters \cite{SchLeiWhi10}.

The goal of the optimisation was to create a lattice of the desired geometry with the highest possible trap frequencies, while having equal barrier heights to all neighbouring sites, using square or triangular geometries. For both geometries, the pattern consists of bands of magnetised material, with edge patterns creating the trapping potentials. Around the trap positions, the boundary of the pattern can be seen as similar to a Z-wire shape. Fig.~\ref{square1d}(a) shows the result for a square lattice. 

The analysis was restricted to single-layer magnetization patterns with binary ``step-like'' magnetization or thickness variations.  This is also how the film is most easily patterned using the lithographic techniques we have at hand.

The optimization algorithm is an adaptation of those used to generate electrode patterns for ion trapping \cite{SchWesLei09}.  Although there are some critical differences between the two systems on a practical level, mathematically the analogies are close.

The film of magnetised material can be written as an integral over dipoles. We describe an infinite periodic lattice of magnetic film with out-of-plane magnetisation $M(\boldsymbol{\rho})$, with
$\boldsymbol{\rho} = (x,y)$ the spatial coordinate, by a Fourier series.  The magnetic potential is then
written as:

\begin{equation}
\label{eq:fourierexpansion}
\begin{split}
\phi(\boldsymbol{r}) = \frac{1}{2}h M_0 \sum_{n,m} e^{-k_{nm}z} \left[C_{nm}\cos(\boldsymbol{k}_{nm}\cdot\boldsymbol{\rho}) + \right. \\ 
+ \left. S_{nm}\sin(\boldsymbol{k}_{nm}\cdot\boldsymbol{\rho})\right],
\end{split}
\end{equation}
where $M_0$ is the value of the magnetization, $h$ is the height of the film, and $\boldsymbol{k}_{nm} = n\boldsymbol{K}_{1} + m\boldsymbol{K}_{2}$ are the reciprocal lattice vectors. 

The lattice field is calculated as  ${\bm B}({\bm r})={\bm B}_{\rm ext}-\nabla \phi({\bm r})$, where ${\bm B}_{\rm ext}$ is a uniform external bias field. The Fourier series can in practice be truncated, keeping only terms with coefficients above some threshold. In general, Fourier components with wavelength short compared to the desired atom-to-chip distance have little effect on the field at the atoms, and can be ignored. The bias field still adds tunability to the final magnetic field and thus the potential energy landscape. We previously explored this feature in the shift register experiment \cite{WhiGerSpr09}. 

In Fig.~\ref{square1d}(b) we show the distribution of the magnetic field $|{\bm B}({\bm r})|$ using a bias field chosen to yield field minima at $z=100$~nm above the surface with equal barriers in the $x$ and $y$ directions. The required bias field is $B_{\rm ext}=(-0.98, -0.39, 0.00)$~mT, the minimum field is $B_{\rm IP}=0.76$~mT, and the trapping frequencies $\omega/2\pi=(2.1, 2.0, 0.8)$~MHz. The trap depth is calculated as 0.29~mT and the barriers 1.3~mT. For the Hubbard model parameters we find $U/J=72$, so that the ground state is expected to be deep in Mott insulator regime.

The bias field can also be tuned to yield low barriers in the $y$ direction, as shown in Fig.~\ref{square1d}(c). The field minima are again at $z=100$~nm. The bias field that achieves this situation is $B_{\rm ext}=(-1.99, -0.04, 0.00)$~mT, the minimum field is $B_{\rm IP}=1.83$~mT, and the trapping frequencies $\omega/2\pi=(1.8, 1.2, 0.6)$~MHz. The trap depth is now 
0.16~mT and the barriers 0.023~mT. Remarkably the number of field minima has been doubled. With a period of 50~nm in the $y$ direction the lattice recoil is now four times higher. We now find 
$U/J=1.4$, well below the Mott insulator transition. 

The pattern in Fig.~\ref{square1d}(a) can be produced in FePt using conventional microfabrication techniques including e-beam lithography followed by reactive ion etching.
The effect of finite resolution is qualitatively equivalent to truncating the Fourier series expansion of the magnetic scalar potential.  Because the contributions of higher Fourier modes (corresponding to small features) on the magnetic pseudo-potential decay rapidly with distance from the surface (Eq.~\eqref{eq:fourierexpansion}), the shape of the potential at the trap position is relatively insensitive to the fine details of the magnetization pattern.

Ultimately, magnetic-film atom chips will be subject to similar physical limitations that limit the density of magnetic data storage, as given by the super-paramagnetic limit.  For small enough feature size thermal fluctuations will drive magnetic domains across the energy barrier associated with reversing the magnetization.

\subsection{Van der Waals attraction to the surface}

At the length scale of hundreds of nanometers, atomic-solid state interactions become particularly important as the trapped atoms approach ever more closely the surface of the atom chip.  The attractive Van der Waals potential shifts the trap minima towards the surface and lowers the trap frequency.  This has to be compensated for by increasing the original magnetic trapping potential.  However the steepness of the trap is limited by requirement that it be slower than the Larmor precession frequency, as otherwise trap loss through Majorana spin flips can occur.

Near the surface the attractive Van der Waals (VdW) potential varies rapidly with distance as $-C_{3}/z^{3}$. Thus the stiffness of the magnetic traps must be large enough to overcome the VdW force. We estimate that at a distance of 100 nm the trap frequency must be greater than approximately 500 kHz. Such high frequencies are unprecedented for magnetic traps, but in our case they will in fact occur quite naturally because the magnetic field gradients close to the surface are extremely high. The atoms are then confined on a very small length scale: for 500 kHz the harmonic oscillator length $\sqrt{\hbar/2m\omega} = 10.7$~nm, or only twice the $s$-wave scattering length. These comparable length scales can give rise to interesting new physics, such as confinement-induced resonances \cite{HalMarNag10}.

We consider atoms trapped very close to a surface, $z\lesssim \lambdabar\approx 124$~nm, therefore we will neglect retardation effects. The potential can be modeled as a harmonic trap plus the Van der Waals potential, 
\beq
	V(z)=\frac{1}{2}m \omega^2 (z-z_0)^2-C_3/z^3.
\eeq
The attractive VdW potential shifts the potential minimum towards the surface and lowers the trap frequency. The potential minimum occurs at $z_t$ where $V'(z_t)=0$. This cannot be solved analytically but we can obtain a lowest order approximation, linear in $C_3$. We write $z_t=z_t(C_3)$, so $z_t(C_3=0)=z_0$ and calculate the lowest order in the power series: $z_t=z_0+C_3 z_t'(0)$. Taking the derivative of the equation $V'(z_t)=0$ with respect to $C_3$ and solving for $z_t'(C_3)$ we get
\beq
	z_t'(C_3)=-\frac{3z_t(C_3)}{m\omega^2 z_t(C_3)^5 - 12C_3},
\eeq
so that the shift in trap position is 
\beq
	\delta z_t=C_3\; z_t'(0)=-\frac{3C_3}{m\omega^2 z_0^4}. \label{trapshift}
\eeq
We want the shift to be small compared to $z_0$, which can be expressed as a condition on the trap frequency,
\beq
	\omega\gg\sqrt{\frac{3C_3}{m z_0^5}} \label{omegacrit_1}
\eeq

Instead of requiring the shift in trap position to be small we can also require that the trap frequency remains positive, i.e. $V''(z_t)>0$. 
\beq
	V''(z_0)=m\,\omega^2-12\frac{C_3}{z_0^5}
\eeq
For this to be positive we require $\omega > 2\sqrt{3C_3/m z_0^5}$, a factor of 2 stricter than Eq.~\eqref{omegacrit_1}. A slightly improved estimate is obtained by expanding $V''(z_0+\delta z_t)\approx V''(z_0)+\delta z_t\;V'''(z_0)$. This leads to the even slightly tighter requirement 
\beq
	\omega > \omega_{\rm crit} = \sqrt{2(1 + \sqrt{6})}\sqrt{\frac{3C_3}{m z_0^5}} \approx 2.63 \sqrt{\frac{3C_3}{m z_0^5}}.
\eeq

In order to calculate numerical values we need the value of $C_3$ which is given by \cite{LanCouAsp96}
\beq
	C_3=\frac{3}{16}\frac{\varepsilon_r-1}{\varepsilon_r+1}\lambdabar^3 \hbar\Gamma.
\eeq
Here $\varepsilon_r$ is the relative dielectric constant of the surface, $\lambdabar=124$~nm and $\Gamma/2\pi=6$~MHz are the reduced wavelength and natural linewidth of the dominant transition from the ground state (for Rb). For metals or high-index dielectrics such as silicon, the factor $(\varepsilon_r-1)/(\varepsilon_r+1)$ is approximately one. Taking the value 0.85 for silicon, we get $C_3=1.3\times 10^{-48}\;$J~m$^3$.
For $z_0=100$~nm we get $\omega_{\rm crit}/2\pi=688$~kHz near a silicon surface.
At the same time the trap frequency $\omega$ must remain small compared to the Larmor (spin precession) frequency, i.e.\ $\omega_{\rm crit}<\omega\ll\omega_L$.


\subsection{Tunneling into the surface}

The small atom-surface distance of 100~nm also raises the question about tunneling of trapped atoms into the surface. Although the magnetic field provides a potential barrier, the attractive Van der Waals potential limits the height of this barrier by 
adding  a diverging attractive $-1/z^3$ core to the trapping potential. This could cause trap losses due to tunneling into the surface. The potential energy is 
\beq
	V(\bm{r})=g_F m_F \mu_B B(\bm{r})-\frac{C_3}{z^3}.
\eeq
The transmission coefficient for tunneling through this barrier can be estimated in the WKB approximation as 
\beq
	T=\exp\left(-2\int_{V(z)>E}\sqrt{\frac{2m}{\hbar^2}[V(z)-E]}\;dz \right).
\eeq
The tunneling rate is exponentially suppressed with barrier height and the magnetic field rises rapidly close to the surface, so that the barrier to the surface quickly becomes much higher than the barrier between neighbors.  

Taking $E\ll V(z)\approx\mu_B B$ we can define a characteristic length scale over which $T$ drops by a factor $e$ as $\ell\equiv\kappa^{-1}=\hbar/\sqrt{8m\,\mu_B\,B}\propto 1/\sqrt{B}$. The proportionality constant is $\hbar/\sqrt{8m\,\mu_B}\approx 1$~nm$\,\sqrt{\rm mT}$. 
For a typical case tunneling to the surface therefore appears to be negligible. For example if we take a 25~nm thick FePt film, patterned with a 100~nm period lattice, and trapping positions at 80~nm from the surface, the above expression yields $1.5\times 10^{-163}$.

\begin{figure}%
\includegraphics[width=\columnwidth]{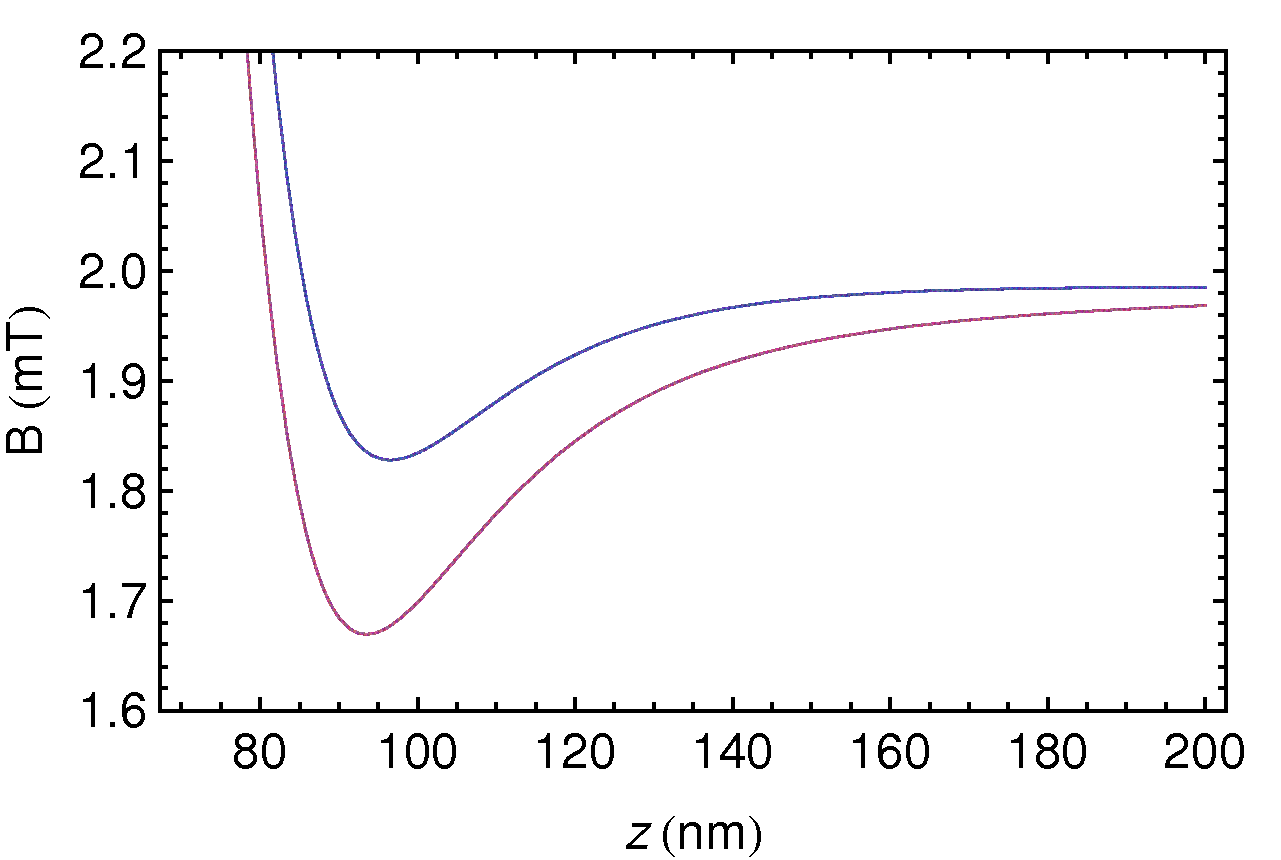}%
\caption{Potential perpendicular to the surface, expressed in magnetic field units, for the situation of Fig.~\ref{square1d}(c). For the upper curve the Van der Waals has been neglected, for the lower curve it has been included.  }%
\label{microK}%
\end{figure}

\begin{figure}%
\includegraphics[width=\columnwidth]{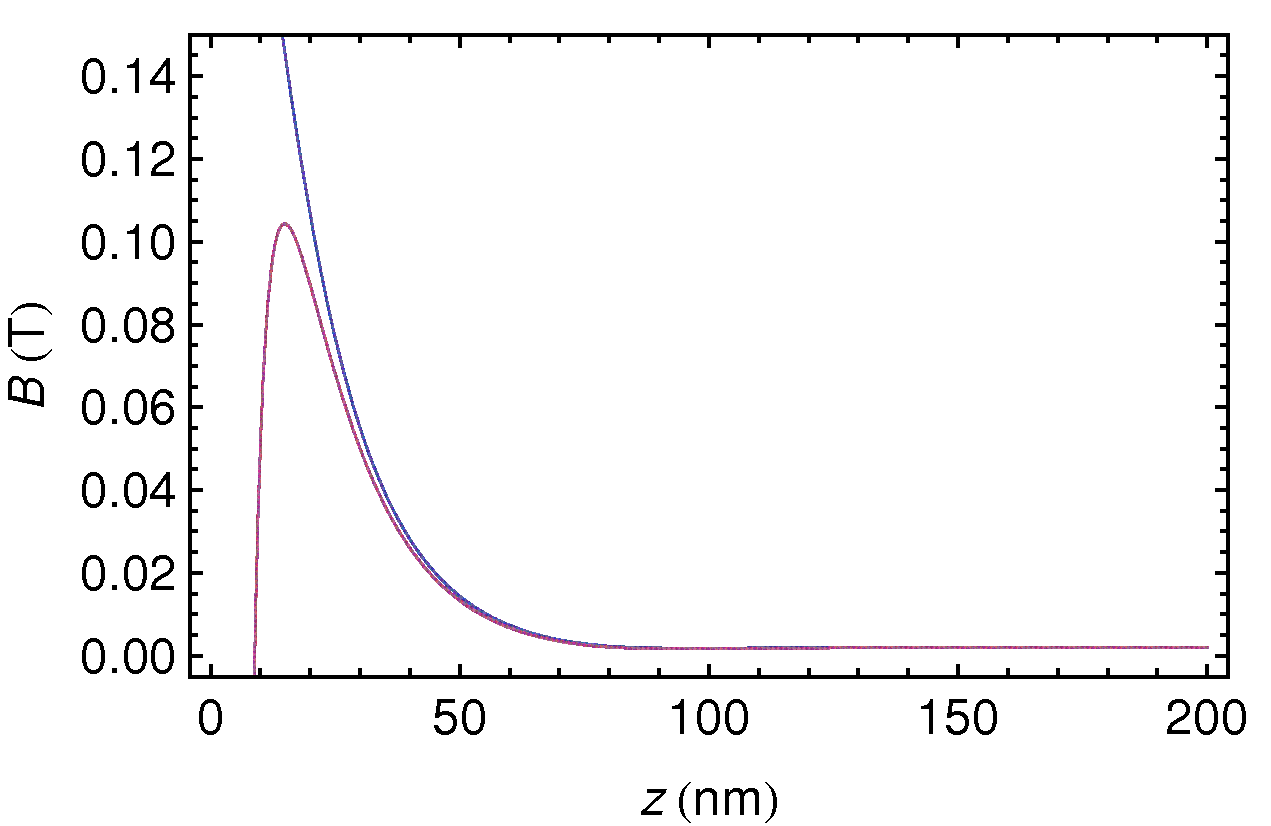}%
\caption{Zoom out of Fig.~\ref{microK}, showing more clearly the potential barrier between the trap position and the trap surface. }%
\label{milliK}%
\end{figure}

\subsection{Johnson noise}

A well known cause of trap loss is Johnson noise in the proximity of a conducting
surface \cite{HenPotWil99,HenPot01}. 
In the vicinity of a conducting material, thermal currents cause magnetic field noise, which can flip the spins of trapped atoms.  The effect is sometimes also called near-field blackbody radiation \cite{HenPotWil99,HenPot01,SchRekHin05,LinTepVul04}.

Using much thinner films than usual in atom chips experiments, we will benefit from the fact that the film thickness is much less than the skin depth at the relevant frequency for spin flips (in the MHz range). The skin depth is $\delta=\sqrt{2/\mu_0 \omega\sigma}$, with $\omega$ the spin flip transition frequency and $\sigma$ the conductivity of the metal; for gold: $\sigma_{\rm Au}=45\times 10^6$~S/m. At a 5~G bias field, for example, we have $\omega=2\pi\times3.5$~MHz, and thus $\delta\approx 40\;\mu$m.  For lattice periods
of 100 nm a magnetic film of only 20 nm thickness will be adequate. Since our atom chips are typically coated by a 50-nm reflective gold coating, this will be the dominant source of Johnson noise induced trap loss. We can obtain an estimate by scaling earlier experimental results \cite{LinTepVul04} or based on theoretical work \cite{SchRekHin05}. 
The first yields 
\beq
	\Gamma=\left(4+\frac{8}{3}\right)^{-1}C_0\left[d(1+d/t)\right]^{-1}.
\eeq
Here $C_0\approx 88\;\mu$s (for Cu), $d=100$~nm is the distance, and $t=50$~nm is the film thickness. 
This yields $\Gamma^{-1}=0.023$~s. 

The lifetime using expressions in Ref.~\cite{SchRekHin05} can also be estimated: 
\beq
	\tau=\left(\frac{8}{3}\right)^2\frac{3\times 10^{22}}{1.7\times 10^6}\left(\frac{\omega}{c}\right)^3 \frac{\delta^2 d^2}{h^2} = 16\;\mbox{ms},
\eeq 
where a film thickness $h=50$~nm and distance from the film $d=100$~nm were inserted. The two estimates are in reasonable agreement with each other. 

These results indicate that we may still benefit from using lower conductivity  material for the conductive coating. Another possibility is to avoid the reflective coating at the position of the lattice. Atoms could first be trapped at a different region of the chip and transported to the lattice either magnetically or optically. As for the magnetic film, dielectric magnetic materials 
would be an interesting option to explore in the future. 

\subsection{Sub-optical detection with tailored electric fields}

The experimental achievement of single-site resolution for ultracold atoms in trapped {\em optical} lattices is currently under active development \cite{BakGilGre09,SheWeiKuh10}. These experiments use optics with very high numerical aperture (NA) to reach sufficient optical imaging resolution in order to extract site-resolved information. A great challenge in these techniques is the required relative positioning stability (at the 10-nm scale) of the lattice sites and the imaging optics.

When atom chips are used to define the sites, our scenario to use electric fields (Sec.~\ref{electricfields}) for site-specific detection via field ionization emerges as an attractive and promising alternative that circumvents the above complications.  

Furthermore, such tailoring of electric fields should be compatible with scaling to the 100-nm or  sub-wavelength regime described in this section.  For instance there is a great interest in the use of carbon nanotubes for field emission (of electrons from the tip) and field ionization (of atoms in vacuum near the tip).  Experiments demonstrating field ionization of cold Rb atoms near carbon nanotubes have recently been reported \cite{GruJagFor09,GooRisHau10}, showing explicitly that the low ionization energy of ground-state alkalis allows for field ionization at fields that are substantially below what is necessary for field emission of electrons from the surface.  The use of nanotubes is very attractive from a conceptual point of view, and may be necessary when scaling down to the dimensions considered here.

\section{Summary and conclusion }

After briefly reviewing our previous result, we have investigated the possibilities to develop a novel, scalable quantum information platform based on arrays of magnetic microtraps defined on a magnetic film atom chip. Two approaches have been presented, characterized by the lattice parameter. 

The first option, with a lattice parameter of 5~$\mu$m is a promising system as a lattice of ensemble qubits. The interactions can be switchable through the transient excitation to Rydberg levels. First measurements on the interaction of Rydberg atoms with the surface have revealed the presence of electric fields due to adsorbed Rb atoms. Some options to mitigate the effects of these adatom fields have been discussed. 

The second option, with a sub-optical lattice parameter of 100 nm, would open up the physics of atomic lattices in new parameter regimes. We estimated the corresponding energy scales in the Hubbard model, which are one to two orders of magnitude higher compared to optical lattices. 
We investigated the conditions to keep Van der Waals attraction to the surface under control and showed that tunneling into the surface will be essentially zero for practical cases.

\acknowledgments

We wish to thank H.B. van Linden van den Heuvell and R. Schmied for fruitful discussions, and C. Ockeloen, S. Whitlock, and R. Thijssen additionally for their contributions to the experiments. This work is part of the research programme
of the ``Stichting voor Fundamenteel Onderzoek der Materie
(FOM)'', which is financially supported by the ``Nederlandse
Organisatie voor Wetenschappelijk Onderzoek (NWO).''




\end{document}